# Tunable frequency band-gap and pulse propagation in a strongly nonlinear diatomic chain

E. B. Herbold[1], J. Kim[2], V. F. Nesterenko[1,2]*, S. Wang[2], C. Daraio[3]

[1] Department of Mechanical and Aerospace Engineering, University of California at San Diego, La Jolla CA 92093-0418 USA

[2] Materials Science and Engineering Program, University of California at San Diego, La Jolla CA 92093-0418 USA

[3] Graduate Aeronautical Laboratories (GALCIT) and Department of Applied Physics, California Institute of Technology, Pasadena, California 91125, USA

One-dimensional nonlinear phononic crystals have been assembled from periodic diatomic chains of stainless steel cylinders alternated with Polytetrafluoroethylene (PTFE) spheres. We report the presence of acoustic band gaps in the dispersion relation of the linearized systems and study the transformation of single and multiple pulses in linear, nonlinear and strongly nonlinear regimes with numerical calculations and experiments. The limiting frequencies of the band gap are within the audible frequency range (20~20,000 Hz) and can be tuned by varying the particle's material properties, mass and initial compression. Pulses rapidly transform within very short distances from the impacted end due to the influence of the band gap in the linear and in nonlinear elastic chain. The effects of an *in situ* band gap created by a mean dynamic compression are observed in the strongly nonlinear wave regime.

# I. INTRODUCTION

The study of pulse propagation in a one dimensional diatomic chain as well as multilayered and periodic samples with a linear interaction law between masses has been a topic of growing interest. This is because the frequency band gaps affect the behavior of the system by prohibiting the propagation of acoustic waves in this part of the frequency spectrum [1-14]. Previous theoretical investigations of linear and weakly nonlinear diatomic systems of macroscopic particles have focused on frequency band-gaps [1-5], localized modes [6, 7] or discrete gap breathers [8, 9] in discrete dynamic systems with various interaction laws between particles.

Investigations of band gaps in discrete metamaterials vary widely in experimental applications. For example, band gaps and strongly localized modes are present in transversely loaded strings with masses [6, 10] and diatomic chains composed of welded spheres (with band gaps in the frequency range of 55-75 kHz) [12]. Layered structures composed from two disparate materials exhibit acoustic band gaps and pass-bands that may be engineered to be very narrow or broad by introducing multiple periods into the layered structure [13]. Pass bands are utilized in non-mechanical systems using a slow-wave microelectromechanical delay line in a chain of coupled resonators [14]. This capacitively-coupled MEMS delay line structure has a band-pass response and signal group velocity as low as 10 m/s, which can be used as a filter in wireless RF or communication systems.

One dimensional chains of elements interacting according to Hertz contact law are appropriate models to investigate linear, weakly nonlinear and strongly nonlinear wave dynamics [15-25]. Their versatility derives from the 'tunability' the wave propagation regime, from linear to strongly nonlinear through an initial static compression. This *a priori* adjustability of acoustic band gaps in strongly nonlinear diatomic systems [26] allows for the development of various practical applications, ranging from acoustic filters, noise mitigation and absorption layers to vibration insulation and the phonon wave guides.

Strongly nonlinear solitary waves (the main mode of signal propagation in weakly compressed chains) exhibit qualitatively different features than weakly nonlinear systems. They have a finite width [15-17, 19-21] that is independent of the solitary wave amplitude, and a pulse speed that is much smaller than the bulk sound speed of the spheres composing the chains [20]. The speed of strongly nonlinear solitary waves observed in assembled phononic crystals can be below the range of sound speed of fluids and gases, as it was demonstrated for PTFE and stainless steel based granular systems [19-24]. The strongly nonlinear wave equations derived for uniform and heterogeneous diatomic chains [15, 20, 26] represent the most general description of wave dynamics including the weakly nonlinear and linear cases.

It should be mentioned that self-demodulation of nonlinear pulses occur in chains of identical spherical particles as well [27]. This occurs when the propagating signal contains frequency components above the cutoff frequency, which depends on the static compression force [28].

This work investigates the formation and propagation of nonstationary signals (quasiharmonic, solitary and shock waves) in linear, weakly nonlinear and strongly nonlinear diatomic periodic chains of particles. The influence of band gaps on signal propagation and their tunability within the audible frequency range are investigated using chains composed of PTFE spheres and stainless steel cylinders.

## II. EXPERIMENTAL PROCEDURES

In these experiments the diatomic chain consisted of a periodic arrangement of PTFE spheres and stainless steel cylinders where the masses were placed vertically in a PTFE cylinder with an inner diameter of 5 mm (shown in Fig. 1 (a)). The chain consisted of 19 PTFE spheres (McMaster-Carr) with diameter $d_s = 2R = 4.76$ mm and mass $m = 0.1226$ g and 18 stainless steel cylinders (McMaster-Carr) with height $h = 3.12$ mm, diameter $d_c = 4.96$ mm and mass $M = 0.485$ g. Stainless steel cylinders were used to ensure a planar geometry of the deformed contact, as assumed for Hertz contact. An additional ferromagnetic steel particle was placed on the top of the chain to magnetically induce a compression force $F_0 = 2.38$ N which decreased the distance between the particle centers by $\delta_o = 7.34$ µm [24].

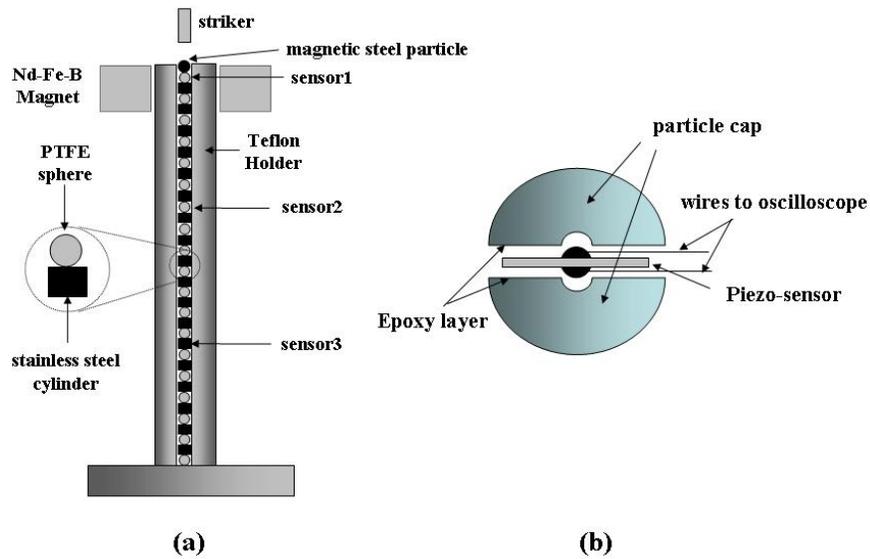

FIG. 1. Schematic diagram of the experimental setup. (a) The diatomic chain is composed

of an alumina striker, stainless steel cylinders, PTFE spheres, embedded sensors (two inside PTFE particles (sensors 1,2) and one (sensor 3) inside stainless steel cylinder) and a Neodymium-Iron-Boron ring magnet for initial static compression. The inset in (a) shows the composition of a unit "cell" as the basis of the diatomic system. (b) Schematic of a PTFE particle with an embedded piezoelectric sensor.

Three calibrated piezoelectric sensors (RC ~ 1 ms) were embedded in two of the PTFE spheres and one stainless steel cylinder and connected to an oscilloscope (Tektronix TDS 2014) to measure the force amplitude and determine the speed of the signals. Each sensor was assembled using lead zirconate-titanate piezoelement (3 mm square plates with thickness 0.5 mm) with nickel-plated electrodes and microminiature wiring embedded in the particles (Fig. 1 (b)). The particles with these sensors were placed in the $2^{nd}$, $14^{th}$ and $27^{th}$ positions from the top (see Fig. 1 (a)). The total mass of each particle including the sensor was approximately equal to the mass of the PTFE particle and stainless steel cylinder respectively. The small mass difference (< 6%) between the particles and particles with sensors were taken into account in the numerical calculations and created negligible effects on the wave propagation. The sensors were calibrated using linear momentum conservation law in separate impact experiments.

The pulses were created by impacts of four different strikers made from $Al_2O_3$ cylinders of different masses, comparable or larger than the mass of one "cell" in the chain. Each cell consisted of one PTFE sphere and one steel cylinder and had a mass of 0.608 g (see the inset in Fig. 1 (a)). The strikers used for the creation of a single and multiple solitary waves were: 0.61 g (about 1 cell mass), 1.22 g (2 cell mass) and 2.75 g (4.5 cell mass) and for shock-like pulses, 17.81 g (about 29 cells mass). The impact velocities used in experiments were 0.44 m/s for the first three masses and 0.20 m/s for the 17.81 g striker.

### III. BAND GAP IN LINEAR ELASTIC DIATOMIC CHAIN

A linear elastic diatomic chain can be assembled from periodic arrangements of elements of different masses and material properties, which are initially compressed by a static force. From an experimental point of view it is convenient to employ the elastic contact of soft spherical particles (PTFE spheres) and elastically rigid cylinders (stainless steel cylinders). The contact between them results in a relatively low elastic modulus of the system leading to a frequency gap within the audible range for possible practical applications.

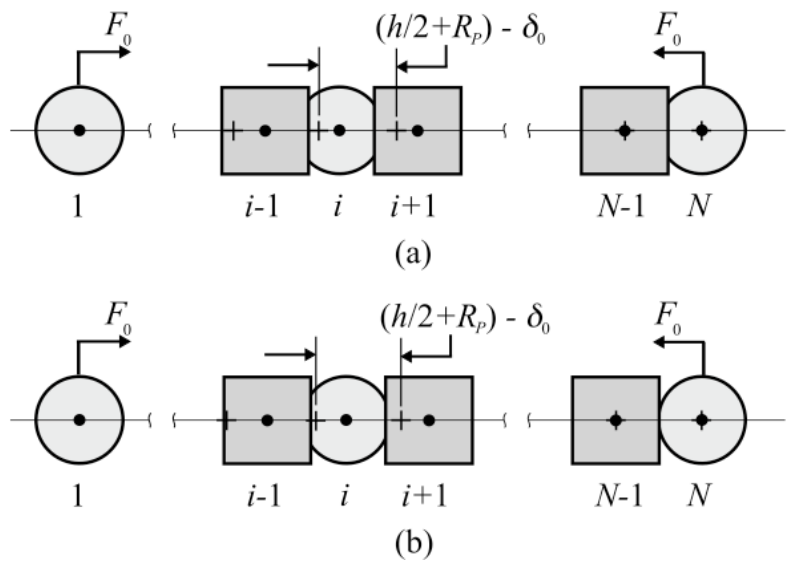

FIG. 2. Relative positions of particles in diatomic mechanical system corresponding to two

wave regimes: (a) strong static compression, (weakly nonlinear case) and (b) weak static compression (strongly nonlinear case). The relative contact area between the particles in the middle of the chain, disturbed by a propagating wave, in comparison with those in undisturbed ends indicate different regimes of behavior.

The diatomic mechanical system studied (PTFE spheres, stainless steel cylinders) is shown in Fig. 2. The crosses and black circles represent the initial and current positions of the centers of the corresponding masses in a statically compressed chain, $h/2$ is half of the height of the cylinder, $R_P$ is radius of sphere, $\delta_0$ is the relative displacement of the neighboring particles due to static compression. The static force may be applied using magnetically induced compression [24]. The linear case presented in Fig. 2 (a) shows that the compression due to the static force $F_0$ is much larger than the perturbed force in the propagating wave. Conversely, the perturbed force in the nonlinear wave is greater than or comparable to the applied force, $F_0$ (Fig. 2 (b)).

The phononic band gap frequencies in an elastic granular chain (Fig. 2 (a)) can be found through the linearization of the force displacement relationship between the two particles in the unit cell:

$$F \approx A(\delta_0 + \delta_d)^{3/2}, \tag{1}$$

where $F$ is the total compressive force including $F_0$ created by the static compression and the dynamic part of the force $F_d$ is due to the wave disturbance. The constant $A$ depends on the elastic properties and geometry of the contacting particles [20],

$$A = \frac{4 E_P E_S (1/R_S + 1/R_P)^{-1/2}}{3[E_S(1-v_P^2) + E_P(1-v_S^2)]}, \tag{2}$$

where $E_P$ and $E_S$, and $v_P$ and $v_S$ are the elastic moduli and Poisson's ratio for the PTFE and stainless steel and $R_S$ and $R_P$ are the is radii of contact of the cylinder with a planar contact surface (in this case $1/R_S = 0$) and PTFE sphere. From Eq. 2 it is clear that elastic properties of PTFE with very low elastic modulus in comparison with steel dominate the interaction force between particles in the system.

The Taylor expansion may be used to linearize Eq. (1) assuming $\delta_o \gg \delta_d$,

$$F \approx A\delta_o^{3/2} + \frac{3}{2} A\delta_o^{1/2}\delta_d. \tag{3}$$

The first term in Eq. (3) is the static force $F_0$ and the linearized form of dynamic force $F_d$. The two force components are $F_0 = A\delta_o^{3/2}$ and $F_d = \frac{3}{2} A\delta_o^{1/2}\delta_d = \beta\delta_d$, where

$$\beta = \frac{3}{2} A\delta_o^{1/2} = \frac{3}{2} A^{2/3} F_o^{1/3}. \tag{4}$$

The linearized forces between particles result in two linearized equations of motion for the cylinders and spheres with displacements $u_{i+1}$ and $v_i$ from equilibrium positions in compressed chain,

$$M\ddot{u}_{i+1} = \beta(v_{i+1} + v_i - 2u_{i+1}) \tag{5}$$

$$m\ddot{v}_i = \beta(u_{i+1} + u_{i-1} - 2v_i), \tag{6}$$

where $i = 1, 2, \ldots, N$, and $N$ is the number of particles (odd indices correspond to spheres and even to cylinders) and $M$ and $m$ denote the masses of the cylinder and sphere. The amplitudes of motion are obtained by substituting the propagating wave solutions represented by $u_{i+1} = ue^{i(kx_{i+1} - \omega t)}$, and $v_i = ve^{i(kx_i - \omega t)}$ into Eqs. (5) and (6) [11],

$$M\omega^2 u = 2\beta u - 2\beta v \cos(ka/2) \qquad (7)$$

$$m\omega^2 v = 2\beta v - 2\beta u \cos(ka/2), \qquad (8)$$

where $a = 2(h/2 + R_P - \delta_0)$ is the unit cell size. Equations (7) and (8) have nontrivial solutions for amplitudes of vibrating masses when the determinant is equal zero. The dispersion relation is:

$$\omega^2 = \frac{\beta}{mM}\left(m + M \pm \sqrt{m^2 + M^2 + 2mM\cos(ka)}\right) \qquad (9)$$

This dispersion relation for the linear elastic diatomic chain is shown in Fig. 3 for $f = \omega/2\pi$ (the plus and minus sign in Eq. (9) correspond to the optical and acoustic branches). The lower and upper bound of the band-gap for this system are expressed by the following equations [11]:

$$f_1 = \frac{1}{2\pi}\left(\frac{2\beta}{M}\right)^{1/2}, \quad f_2 = \frac{1}{2\pi}\left(\frac{2\beta}{m}\right)^{1/2}. \qquad (10)$$

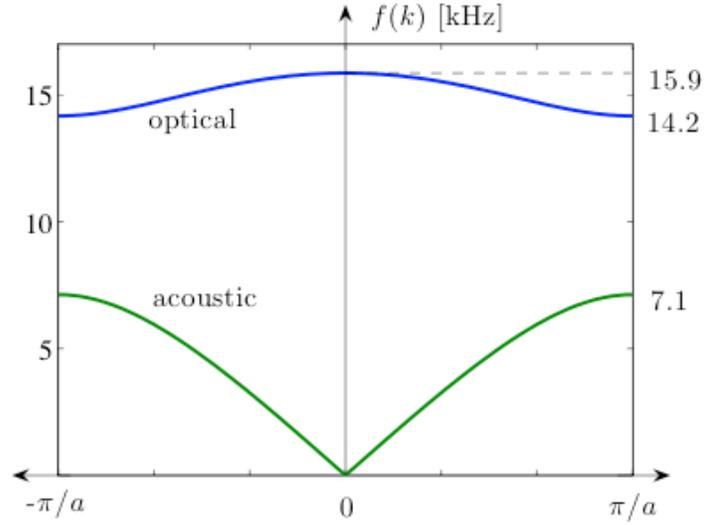

FIG. 3. Dispersion relation in the first Brillouin zone for linear elastic diatomic chain composed of stainless steel cylinders and PTFE spheres used in the experiments.

The band-gap frequencies may be tuned by changing the value of $\beta$ that depends on $A$ and the initial compressive force $F_0$. The band gap is sensitive to the material properties $E$ (Young's modulus), and $v$ (Poisson's ratio), the particle radii $R$ and the initial static compression (Eqs. (2) and (3)). Note that the band gap frequencies are nonlinearly dependent on the initial compression force ($f \sim F_0^{1/6}$).

The corresponding elastic moduli and Poisson's ratio for the diatomic chain are $E_P = 1.46$ GPa [21], $E_S = 193$ GPa, and $v_P = 0.46$ and $v_S = 0.3$, respectively. These parameters based on Eqs. (2), (4) and (10) with $F_0 = 2.38$ N ($\delta_0 = 7.34$ μm) result in the limiting frequencies $f_1 = 7{,}120$ Hz and $f_2 = 14{,}162$ Hz (Fig. 3) that are within the audible range from 20-20,000 Hz.

## IV. RESULTS AND DISCUSSIONS

### A. Signal transformation in a tunable linear elastic diatomic chain

Signals with amplitudes much smaller than the magnetically applied static force propagate in the linear regime. It is important to find the transient response of these signals due to initial conditions that are close to a harmonic excitation with frequencies within the band gap to find a characteristic spatial and temporal point where the band gap is able to affect the shape of the propagating pulse. It should be mentioned that the existence of band gap does not answer the question when entering signal will be affected by it.

In the linear case the pulses are transformed by dispersion [20] and the role of nonlinearity is not essential. In the following figures the pulses and their corresponding frequency spectrum are plotted for comparison. The abscissa values of the frequency plots are the Fourier coefficients derived from a fast-Fourier-transform,

$$C_k = \hat{X}(k)/M = \sum_{j=1}^{N} X(j)/M \exp[-2\pi i(j-1)(k-1)/N], \qquad (11)$$

where $X(j)$ denotes the vector with length $N$ to be transformed, $j$ and $k$ increment over time and frequency and $M$ is the length of $X(j)$ without 'padded' zeros. For example, $C_0$ denotes the mean of the transformed function. In each of the following figures, a ratio of $N/M = 5$ was used since the pulses were finite and not exactly periodic. These values were chosen for ease of correlation between the pulse and its spectrum.

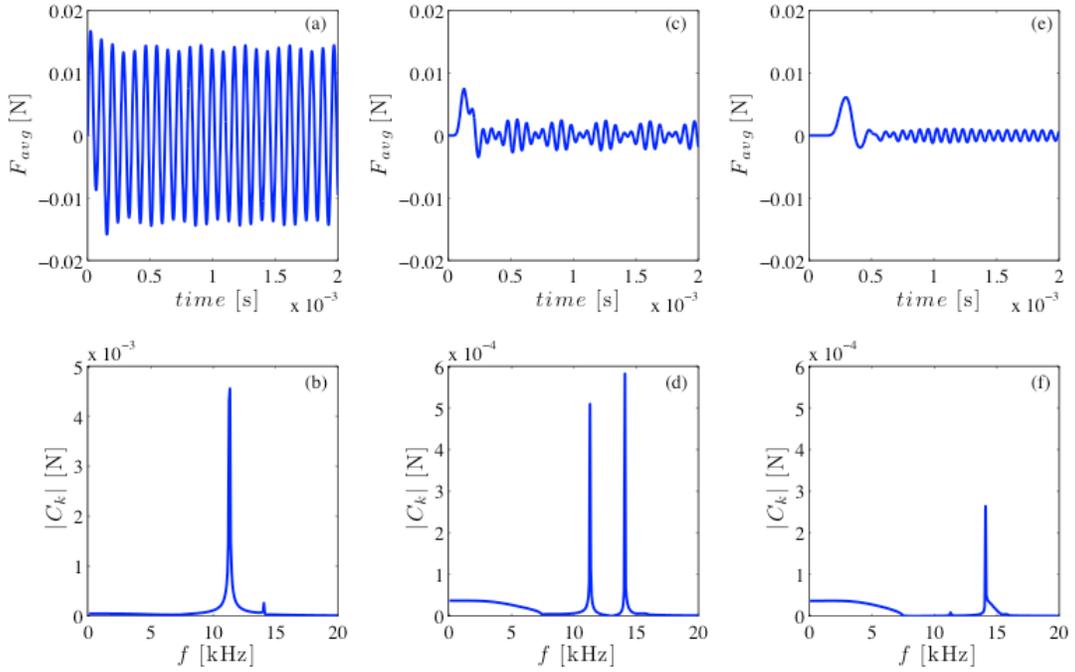

FIG. 4. Linear chain with a static force $F_0 = 2.38$ N. The top PTFE sphere was given an initial velocity of 0.00442 m/s. (a) Dynamic force between the top PTFE particle and stainless steel cylinder and (b) its Fourier spectrum; (c) Force pulse inside 5th particle (PTFE) (averaged of contact forces between $4^{th}$ (SS) and $5^{th}$ (PTFE) and $5^{th}$ (PTFE) and $6^{th}$ (stainless steel cylinder) particles) and (d) its Fourier spectrum; (e) Force in transmitted pulse inside 11th (PTFE) particle and (f) its Fourier spectrum.

In the numerical calculations an oscillatory pulse with a frequency inside the band-gap was created at one end of the periodic diatomic chain (Fig. 1) composed of 300 cells, with a PTFE sphere at the top. This PTFE sphere was compressed by a static force of 2.38 N and given an initial velocity of 0.00442 m/s. This resulted in an oscillatory excitation that was approximately harmonic (Fig. 4 (a) and (b)). The amplitude of the dynamic force between the top PTFE sphere and the stainless steel cylinder was 0.015 N (160 times smaller than initial compression force) and did not allow the top PTFE particle to separate from the chain. The majority of the frequency spectrum in Fig. 4 (b) is between the band-gap frequencies with its largest frequency component at 11 kHz. The Fourier spectrum for all cases were found for propagating pulses up to 15 ms, but the data in Fig. 4 (a), (c) and (e) are truncated to clearly show the leading pulse.

The numerical results of the force in the propagating pulse (i.e. the force inside $5^{th}$ PTFE particle belonging to the third cell and the force inside $11^{th}$ PTFE particle belonging to the sixth cell) and their frequency spectrums are shown in Fig. 4 (c) and (d) and Fig. 4 (e) and (f) respectively. The amplitude of the first pulse at the $6^{th}$ cell is 0.006 N, which is about four hundred times lower than the initial compression force and its width is about nine particles (or 4.6 cells). At the $11^{th}$ particle the wavelength increases to 4.7 cells. At the $21^{st}$ particle wavelength of the propagating pulse becomes almost twelve particles (or about 5.9 cells). It is interesting that the initial wavelength is close to 5 cells similar to the pulse width in a strongly nonlinear diatomic chain without initial compression [20, 24].

It is clear from Fig. 4 that this linear chain is able to dramatically change the shape of the initial oscillatory pulse within relatively short distances from the top PTFE particle (within only 6 cells!), which is mainly due to dispersion. As the signal propagates, the frequency spectrums in Fig. 4 (b), (d) and (f) show that the pulse preferentially transforms (demodulates) such that the components in the band gap move toward lower and higher frequencies.

The Fourier spectrum of the force in the system very close to the top of the PTFE

sphere (inside 5$^{th}$ PTFE particle, Fig. 4 (c) and (d)) demonstrates significant frequency components above and below the band gap in addition to the initial signal. This is significantly different than the spectrum of the excited pulse (Fig. 4 (a) and (b)). There are two main frequencies in Fig. 4 (d): one is the source signal at about 11 kHz and the other is about 14 kHz, which was also present in the input signal at about the same amplitude. These two harmonics are responsible for the beating phenomenon observed in the force signal (Fig. 4 (c)). This beating disappears and the trailing oscillations become approximately harmonic at the 11$^{th}$ particle (PTFE) (Fig. 4 (e)) due to the complete transformation of the 11 kHz frequency component.

It should be mentioned that the high frequency corresponding to the top of the band gap does not disappear during signal propagation. This is due to the oscillatory motion of the light PTFE particle in the cell, which is also a characteristic of nonlinear signals [20, Chap. 1, Fig. 1.19]. The relatively small peak around 14 kHz in Fig. 4 (b) is present in the initial signal due to the motion of top PTFE particle against the steel cylinder under constant force. The amplitude of this frequency component decays very slowly (compare Figs. 4 (b), (d) and (f)) in striking contrast to transformation of the initial 11 kHz component.

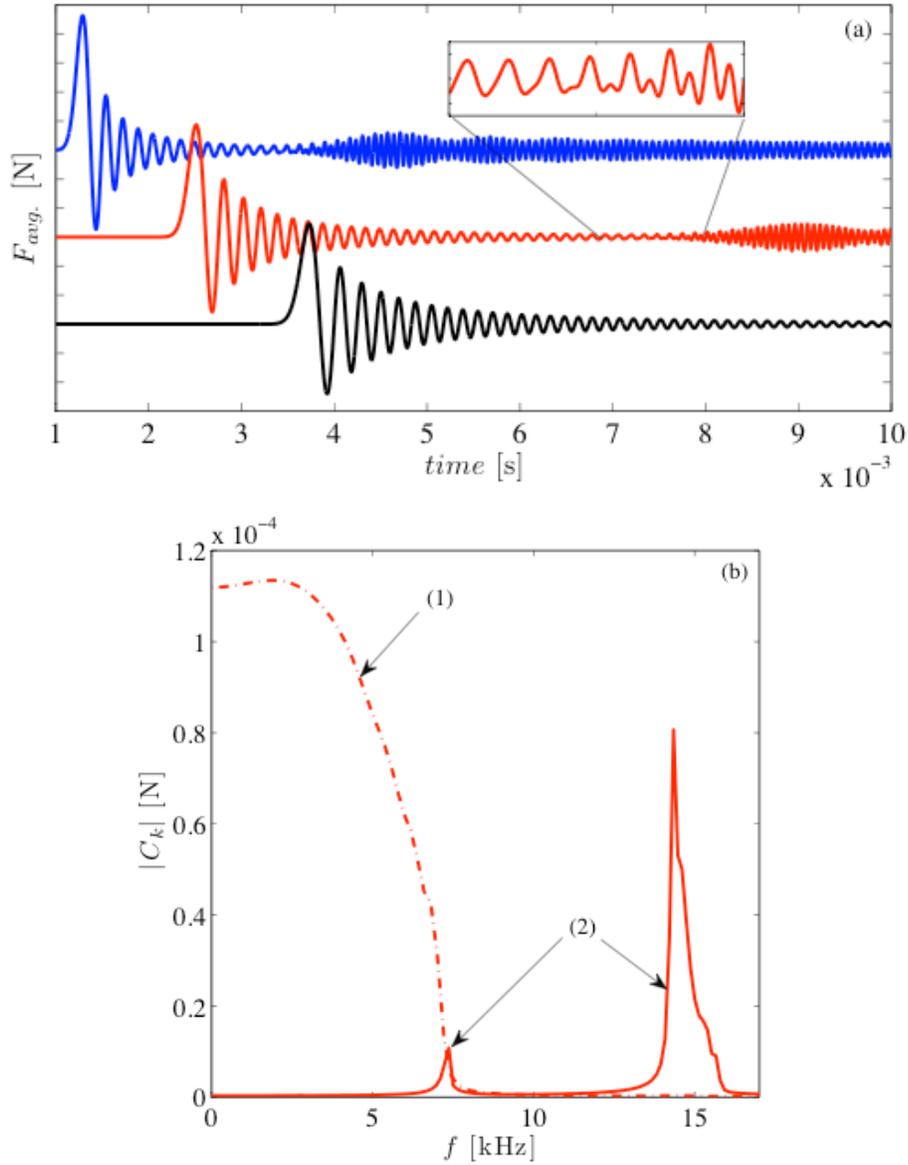

FIG. 5. Linear chain, static force $F_0 = 2.38$ N, the top PTFE sphere was given an initial velocity of 0.00442 m/s. (a) Forces in the transmitted pulses inside 50$^{th}$ particle (PTFE) (averaged of contact forces between 50$^{th}$ (stainless steel cylinder) and 51$^{st}$ (PTFE) and 51$^{st}$ (PTFE) and 52$^{nd}$ (stainless steel cylinder) particles, inside the 101$^{st}$ particle and inside 151$^{st}$ particles. The vertical scale is 0.001 N/div. The inset in (a) shows the transition from acoustic to optical modes in the 101$^{st}$ particle. (b) Curve (1): The Fourier spectrum for leading pulse for 101$^{st}$ particle up to the time 7 ms with mainly acoustic frequency and curve (2) is the Fourier spectrum for the trailing wave group up from the time 7 to 12 ms containing the optical frequency component.

It is interesting to investigate pulse transformation at larger distances from the top particle to investigate the behavior of the leading pulse. The shapes of the pulses are presented in Fig. 5 (a) for the 51$^{st}$, 101$^{st}$ and 151$^{st}$ particle. In Fig. 5 (a) the pulse continues to change due to linear dispersion; particularly the leading pulse is not of solitary shape at these distances from the top particle as it look closer to the entrance (Fig. 4 (c),(e)). The leading part of the propagated signal measured at the 101$^{st}$ particle up to 7 ms has a broad band of frequencies below the band gap (curve 1 in Fig. 5 (b)) and after 7 ms the signal is characterized by one peak at the bottom and top of the band gap (curve 2 in Fig. 5(b)). It should be mentioned that the harmonic with frequency about 14 kHz continues to decay with the pulse propagation (compare Figs. 4 (d), (f) with Fig. 5 (b), curve 2). The wave packet containing frequencies at the top of the band gap (not shown on the bottom wave profile in Fig. 5 (a)) propagates with a lower speed in comparison with the leading oscillatory pulse composed of frequencies lower than 7 kHz. It should be mentioned that there is no stationary single pulse in the linear dispersive case and the leading pulse will evolve further during propagation (for example, with decreasing amplitude and increasing width), but the main features of it are already established before it reaches the 100$^{th}$ particle.

### B. Signal transformation in nonlinear elastic diatomic chain

The band gap is a property of linear systems (Eqs. (7) and (8)) and it is not clear if it can influence the signal propagation in the nonlinear regime. Recently, the effects of varying isotropic static compression in the linear and nonlinear regimes for triangular (two-dimensional) lattices of identical steel spheres shows evidence of self-demodulation in the time-frequency plots of acoustic signals [28]. Here, the behavior of signals in diatomic chains with amplitudes comparable to the initial compression is investigated numerically.

A nonlinear pulse in the same diatomic chain as the previous section was created when the top PTFE particle was given a higher initial velocity of 0.442 m/s, which is two orders of magnitude larger than in the linear system (Fig. 6 (a) and (b)), though the particles in the chain remain in contact. This resulted in a similar oscillatory excitation shown in Fig. 4 (a) and (b), but the dynamic force between the top PTFE sphere and stainless steel cylinder have an amplitude comparable to the static compression force (being 1.6 times smaller). The numerical results of the force in the pulse propagating inside the system and their frequency spectrums are shown in Fig. 6 (c) and (d) and Fig. 6 (e) and (f) respectively. The force inside the $5^{th}$ PTFE particle belongs to the third cell and the force inside the $11^{th}$ PTFE particle belong to the sixth cell. The amplitude of the first pulse is at the $6^{th}$ cell is 0.65 N and is about 3.7 times lower than the initial compression force and its width is about nine particles (or 4.6 cells). This pulse is widening when it propagates in the system. For example at the $21^{st}$ particle the width is approximately eleven particles (or about 5.7 cells), which is closer to the width of a stationary pulse in strongly nonlinear diatomic chains without static compression [20, 24].

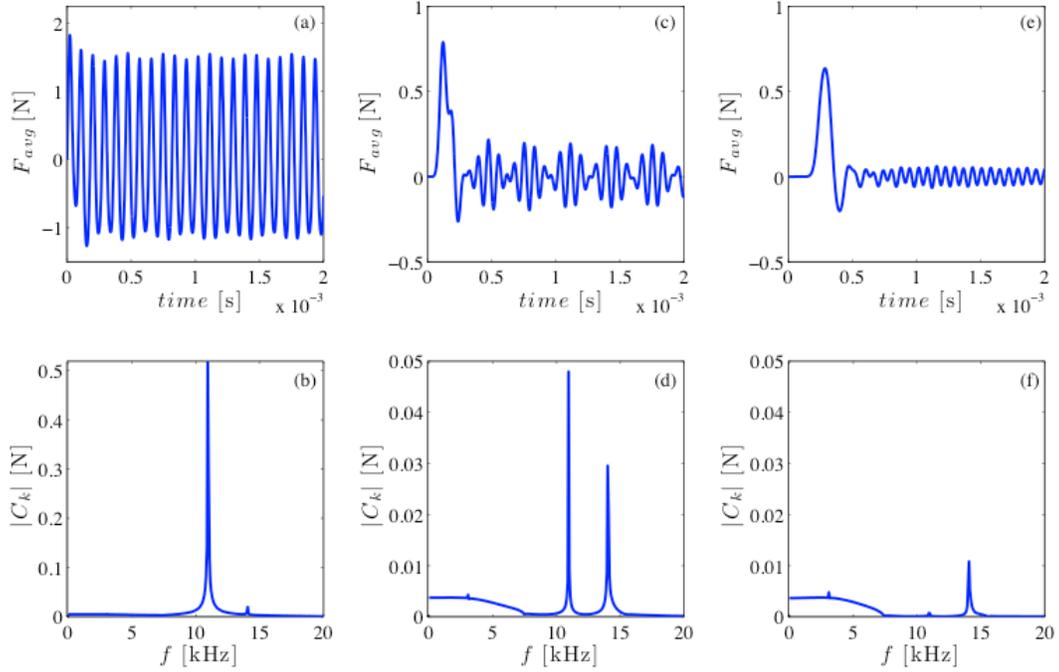

FIG. 6. Nonlinear chain with a static force of $F_0 = 2.38$ N applied to the top PTFE sphere, which was given an initial velocity of 0.442 m/s. (a) The force between the top PTFE particle and stainless steel cylinder (incoming pulse) and (b) its Fourier spectrum; (c) The force in the transmitted pulse inside the 5$^{th}$ particle (PTFE) (averaged of contact forces between 4$^{th}$ (SS) and 5$^{th}$ (PTFE) and 5$^{th}$ (PTFE) and 6$^{th}$ (stainless steel cylinder) particles and (d) its Fourier spectrum; (e) The force in transmitted pulse inside 11th (PTFE) particle and (f) its Fourier spectrum.

At relatively short distances from impacted end (about 6 cells) it is clear that this nonlinear chain dramatically changes the shape of the pulse as its frequency spectrum shifts toward lower and higher frequencies. The behavior of the signal in this nonlinear chain is similar to the case of linear chain, suggesting that the nonlinearity of the chain does not influence the signal propagation as much as dispersion. This means that, for practical applications, the influence of the band gap is also relevant for finite amplitude nonlinear signals within short distances from the signal source.

The pulse transformation as it propagates into the system at larger distances from the impacted end (up to the 151$^{st}$ particle) is shown in Fig. 7. As in the linear case (Fig. 5) the leading propagating pulse continues to transform due to dispersion even in the presence of nonlinearity. However, at much larger distances, the leading solitary wave, which is a steady solution in weakly nonlinear case, may separate from the oscillatory part of the signal in contrast with the linear case.

At the 101$^{st}$ particle the leading oscillating pulse (up to 7 ms) has a broad band of frequencies below the band gap (curve 1, Fig. 7(b)) and for the interval from 7 ms to 12 ms it is characterized by two peaks at the vicinity of the bottom ($f_1$) and one at the top of band gap ($f_2$) (curve 2, Fig. 7(b)). In contrast to Fig. 5(b) corresponding to linear case there is an additional 3 kHz component, which may be due to the influence of nonlinearity in the system. The wave packet containing frequencies above 14 kHz (not shown on bottom wave profile) propagates with a lower speed than the leading oscillatory pulse composed of frequencies lower than 7 kHz. The oscillations of the first (compressed) particle has a harmonic at ~14 kHz (small peak in Fig. 6(b)), which results in signals that propagate into the system with a similar frequency and with a low speed along with the leading pulses that travel with speeds corresponding to the acoustic branch.

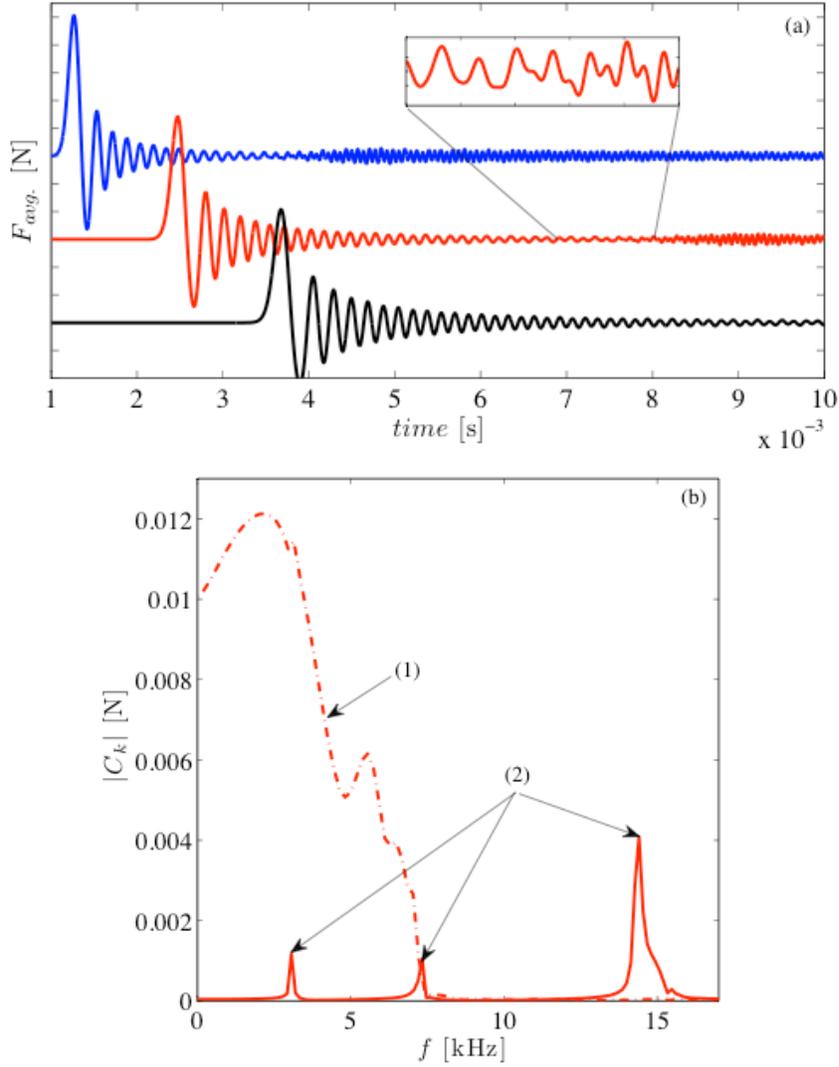

FIG. 7. Nonlinear chain with a static force of $F_0 = 2.38$ N applied to the top PTFE sphere, which was given an initial velocity of 0.442 m/s. (a) Forces in the transmitted pulses inside the 50th particle (PTFE) (averaged of contact forces between $50^{th}$ (stainless steel cylinder) and $51^{st}$ (PTFE) and $51^{th}$ (PTFE) and $52^{nd}$ (stainless steel cylinder) particles, inside $101^{st}$ particle and $151^{st}$ particles. The vertical scale is 0.1 N/div. The inset in (a) shows the transition from acoustic to optical modes in the $101^{st}$ particle. (b) Curve (1): Fourier spectrum for the leading pulse for the $101^{st}$ particle, up to 7 ms containing acoustic

frequencies and curve (2) is the Fourier spectrum for the trailing wave group (taken from 7 ms to 12 ms), containing acoustic and optical frequencies.

Since the gap in the frequency spectrum is a property of a linear elastic system it is surprising that in the nonlinear chain, which is a qualitatively different system (for example, it supports solitary wave absent in a linear case), the signal transformation is similar to linear chain. In the trailing wave group we have no component inside band gap, only components at the top of acoustic and bottom of optical bands. The frequency at the top and the bottom are long living in comparison with those inside band gap in both linear and nonlinear chains. So, waves with finite amplitude in a nonlinear material are sensitive to the same range of band gap frequencies as waves with infinitesimally small amplitudes as in linear chains. This allows tunability of band gap in nonlinear range of material response based on Eq. (14) derived for infinitely small amplitude of signals.

To compare experimental results with the numerical investigation for the nonlinear chain four separate experiments were performed in a diatomic chain composed of 37 elements (including magnetically induced static compression and the gravitational preload) with strikers having different masses and velocities: 0.61 g, 1.22 g, and 2.75 g alumina strikers at $v_0 = 0.44$ m/s and a 17.81 g striker at $v_0 = 0.2$ m/s. Each striker generated incoming pulses of different lengths and amplitudes. In each case, the static compression was 2.38 N, applied to the top magnetic spherical particle (in previous numerical calculations force was applied to the top PTFE particle to generate practically harmonic boundary conditions). The experimental set up is shown in Fig. 1 and experimental data are presented in the Table I. There is a reasonably good agreement between pulse velocities in experiments and numerical calculations despite dissipation present in experiments.

Table 1. Comparison of average speed $V_s$ of leading pulses taken between particles number $14^{th}$ and $27^{th}$ for experiments and numerical results in the nonlinear chain for different masses and velocities of strikers.

|  | Striker Mass [g] | Pulse Speed [m/s] |
|---|---|---|
| Experimental | 0.61 | 175 |
|  | 1.22 | 177 |
|  | 2.75 | 181 |
|  | 17.81 | 179 |
| Numerical | 0.61 | 171 |
|  | 1.22 | 173 |
|  | 2.75 | 176 |
|  | 17.81 | 169 |

The results of the impact by the 0.61 g striker are shown in Fig. 8. The impact resulted in a single pulse at the top of the chain since the striker mass was close to the mass of one cell in the system. The initial signal was quite different (due to the boundary conditions) from Fig. 6 (a) although the impulse in experiments also transformed into a single leading pulse (Fig. 6 (c) and (e)).

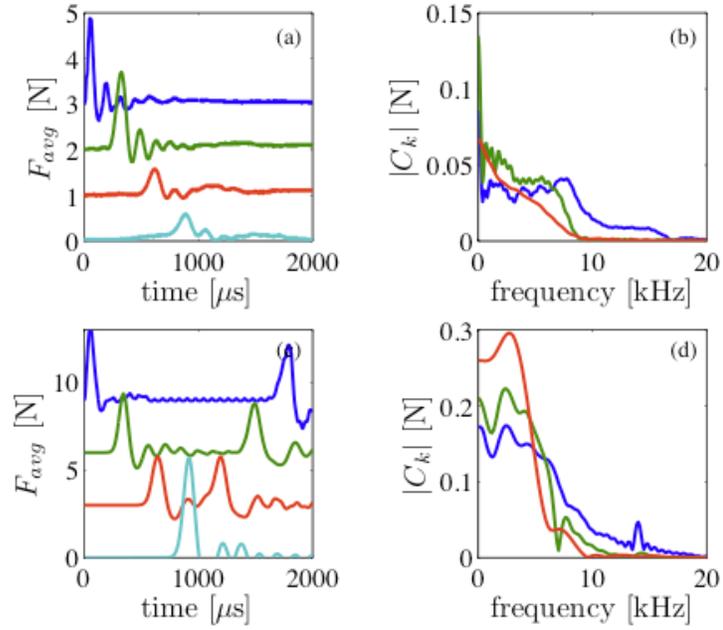

FIG. 8. Nonlinear chain with a static force of $F_0 = 2.38$ N applied to the top magnetic particles impacted by alumina striker with an initial velocity of 0.44 m/s. (a) Dynamic force in experiments recorded in the first PTFE particle (incoming pulse) and in sensors placed in the PTFE particle in the seventh cell and in the steel cylinder in the 13$^{th}$ cell and in the wall respectively; (b) Fourier spectrum for forces presented in (a) (c) Dynamic force in numerical calculations corresponding to experimental conditions in (a); (d) Fourier spectrum for forces presented in (c).

Figure 8 shows the experimental results (Fig. 8 (a)) and the corresponding numerical calculations (Fig. 8 (c)) as well as the Fourier spectrum of each in Figs. 8 (b) and (d). It is clear from Fig. 8 (b) and (d) that the frequency spectrum of the initial disturbance is inside the band gap (7.1 kHz – 14.2 kHz).  It is clear that the signal is shifted toward lower frequencies within 7 cells from the impacted end. The dynamic compression of the chain effectively shifts the lower boundary of band gap toward a higher frequency ~9 kHz than the predicted lower limit of 7.1 kHz corresponding to static precompression (*in situ* modification of band gap) explaining higher frequencies in the propagating pulse in comparison with linear case. This fast modification of the signal is similar to previously observed for practically harmonic incoming excitations in linear and nonlinear cases (Figs. 4 and 6 (b), (d), and (f)). It should be mentioned that similar behavior of Fourier spectrum corresponding to numerical calculations and to experimental data suggests that dissipation (apparently affecting amplitude of signal in experiments) is not the main cause for the observed signal transformation in the frequency domain. A similar behavior of the signals in frequency domain both in experiments and numerical calculations was observed for pulses excited by the impact of an alumina striker (1.22 g at 0.44 m/s) generating a series of pulses.

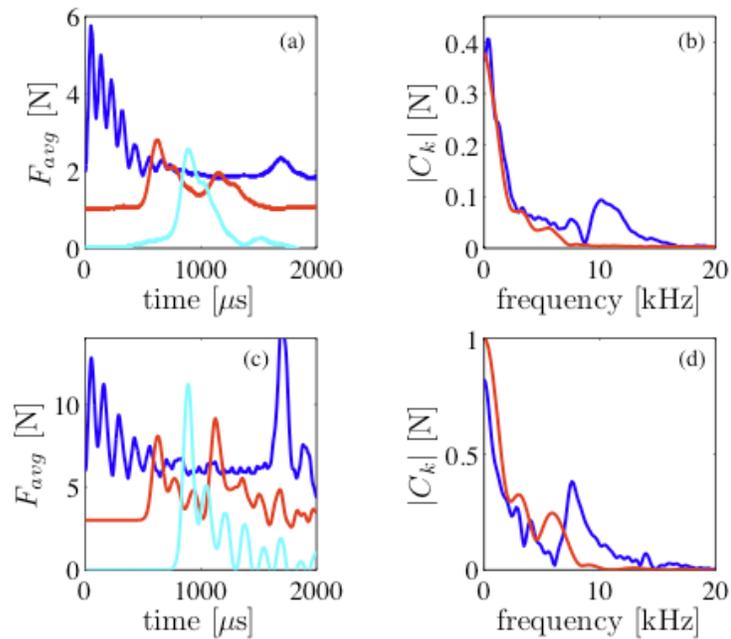

FIG. 9. Nonlinear case, static force $F_0$ = 2.38 N. The striker had a mass equal to 4.5 cells (2.75 g) and an initial velocity of 0.44 m/s. (a) Dynamic force in experiments between first PTFE particle and stainless steel cylinder (incoming pulse) and between sensors in the steel cylinder in the 13$^{th}$ cell and in the wall respectively; (b) Fourier spectrum for forces presented in (a); (c) Dynamic force in numerical calculations corresponding to experimental conditions in (a); (d) Fourier spectrum for forces presented in (c).

To increase the duration and amplitude of the incident signal, experiments were conducted using an alumina striker of larger mass (2.75 g at 0.44 m/s). In Fig. 9 the experimental and numerical results of the impact of this striker are shown. This impact created a sequence of connected pulses in experiments and numerical calculations and the corresponding frequency spectrum of the incoming pulses are shown in Fig. 9 (b) and (d). The frequency spectrum shows that the initial pulse corresponding to the force inside the second (PTFE) particle has a significant portion within the band gap. In comparison with Fig. 8 the main portion of the frequency spectrum within the band gap extends to a slightly higher frequency. This may be due to an additional effect of a larger mean effective compression in addition to the static force and thus, increases the lower limit of band gap (*in situ* modification of band gap). From a comparison of the data for the 2$^{nd}$ and 27$^{th}$ (PTFE) particle it is clear that the spectrum has shifted to lower frequencies, spectrum for the pulse reflected from the wall is not shown for clarity of the graphs. In experiments and numerical calculations the spectrum cut-off observed for the 27$^{th}$ particles is slightly higher than the predicted lower limit of the band gap from the linear approach (7.1 kHz).

A comparison between experimental and numerical data reveal smaller time intervals between peaks in experiments (compare Figs. 9(a) to 9(c)) resulting in a component of Fourier spectrum with higher frequency than in corresponding numerical calculations (compare Figs. 9(b) to 9(d)). We explain this behavior by a delayed separation of solitary pulses in experiments due to dissipation.

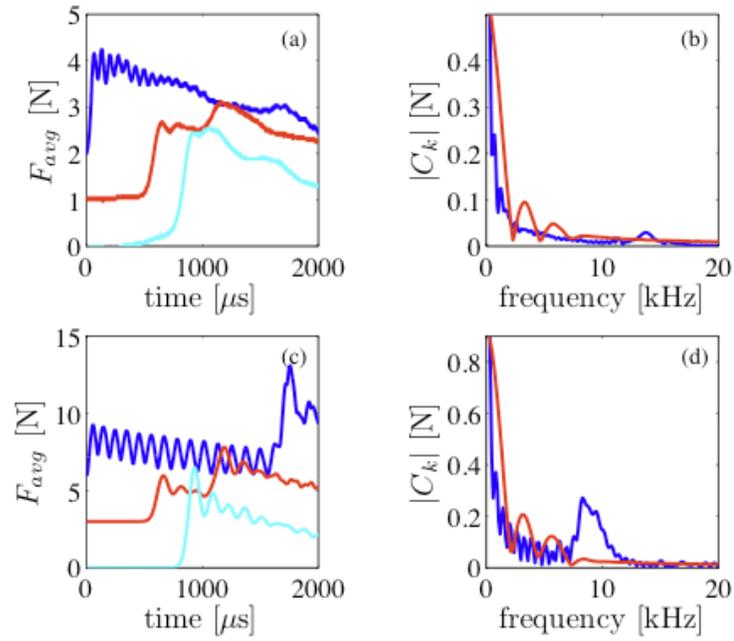

FIG. 10. Oscillatory shock waves excited in the nonlinear case in a diatomic chain composed of 37 elements, static force $F_0 = 2.38$ N. The striker had a mass equal to 17.81 g (about 29 cells mass) and an initial velocity of 0.2 m/s. (a) Dynamic force in experiments in the first PTFE particle and between sensors in the PTFE particle in the 7[th] cell and in the steel cylinder in the 13[th] cell and in the wall respectively; (b) Fourier spectrum for forces presented in (a); (c) Dynamic force in numerical calculations corresponding to experimental conditions in (a); (d) Fourier spectrum for forces presented in (c).

To create an oscillatory shock wave in the experiments we used impact of an alumina striker with a mass of 17.81 g at 0.2 m/s. In Fig. 10 the experimental and numerical results corresponding to the impact of this striker and the corresponding frequency spectrum of the pulses are shown in Fig. 10 (b) and (d) are shown.

The frequency spectrum shows that the initial pulse in the second (PTFE) particle has a significant portion within the band gap. However, optical modes are not excited in this case. To check if this mode appears as signal propagates the numerical calculations were extended up to 15 ms, but the high frequency component (at about 14 kHz) corresponding to the optical mode trailing the leading shock wave was not observed. This contrasts the linear case with practically harmonic boundary conditions evolving into a leading pulse with an oscillatory tail (Fig. 4 (e), (f)). It may be concluded that for all shock loaded cases, the high frequency component is not excited at the boundary.

In comparison to Fig. 9 the main portion of the frequency spectrum within the band gap is centered about a slightly higher frequency in numerical calculations (7.7 kHz in Fig. 9 (d) and 8.4 kHz in Fig. 10(d) correspondingly). This may be due to the effect of a compressive dynamic force creating a larger mean effective precompression resulting in a higher effective stiffness of the compressed contacts in comparison with initial state.

The oscillations in the initial shock in experiments are significantly damped in the 27$^{th}$ (PTFE) particle and the spectrum shifted to lower frequencies. This can be caused by dissipation if it is larger than some critical value [29]. It is interesting and unexpected that this was also the case in the numerical calculations despite the fact that dissipation was not included. We explain the mechanism of the effective smoothing of the oscillating shock profile without dissipation in numerical calculations by the influence of band gap that eliminate the corresponding harmonic in the incoming pulse centered at about 9 kHz within band gap. The effective compression in the shock wave shifts the lower boundary of band gap toward a higher frequency ~9 kHz than the predicted lower limit of 7.1 kHz (*in situ* modification of band gap). Although the initial impact creates an oscillating shock, as in one mass chain [20, 29, 30], these oscillations cannot propagate due to band gap.

Despite the fact that both experimental and numerical incoming profiles were initially oscillating, the frequencies of oscillations were quite different with a higher frequency being observed in experiments (9 and 12 kHz correspondingly). The mechanism of generating higher frequencies in experiments can be due to slower separation of the pulses in presence of dissipation effectively increasing the observed frequency. To check this hypothesis viscous dissipation was included in numerical calculations (similar to [29, 30]) resulting in harmonics with higher frequency (their frequency increasing with viscosity) in comparison to calculations without dissipation. It should be mentioned that magnetorheological liquids allow effective tunability of viscosity with magnetic field [31], which can be used to tune shock response of granular system immersed in such media.

**B. Signal transformation in strongly nonlinear diatomic chain**

If the amplitude of signal is significantly larger than the initial static compression, qualitatively new features can be expected in diatomic chains than in linear and nonlinear pulses. It should be mentioned that in an initially uncompressed chain ("sonic vacuum") there is no characteristic band gap that is representative of a linear elastic system. At the same time the compressed state behind a relatively long "shock wave" can exhibit a phenomena caused by *in situ* band gap where properties depend on the average compression caused by the wave. This was seen in the previous section for nonlinear impulses (Figs. 9 and 10).

The propagation of strongly nonlinear waves in diatomic chains is not directly comparable to the previous two cases since the separation of particles is now possible with the absence of initial compression. This also occurs when the initial impact velocity value is large enough for the particle to overcome the static force. The frequency spectrum of the propagating compressive signal should not be affected by the band gap (Eq. (14)), since both $f_1$ and $f_2$ are very small due to very weak gravitational compression in a vertical chain. In the case of impact by a particle with a mass close to the mass of the cell a single solitary pulse will have a frequency spectrum close to the solution to the long-wave approximation given in [20, 26]. Higher amplitudes of strongly nonlinear solitary waves result in higher frequencies since the spatial size does not depend on amplitude, but the signal speed increases with wave amplitude. Thus, the allowed frequencies of harmonics composing the single solitary wave are tuned by amplitude of wave.

It is interesting to investigate this case because it will elucidate the effect that the length of the impact has on the resulting frequency spectrum. It was seen in the previous nonlinear case that the additional dynamic compression in the pulse may shift the initial frequency spectrum to higher frequencies.

To compare experimental with numerical results for the strongly nonlinear chain four separate experiments with an identical set up as in the previous sections (Fig. 1) were performed with the exception of the magnet used to induce static compression in the previous cases. The only static force in this case is due to gravitational preload in experiments and calculations. Each experiment has incoming pulses of different lengths and amplitudes depending on mass of the striker to excite different Fourier spectrums and observe their evolution. The results of the experiments for a chain of PTFE spheres and steel cylinders in strongly nonlinear regime are shown in Fig. 11-13 and Table 2. There is a reasonably good agreement between pulse velocities in experiments and numerical calculations despite dissipation present in experiments.

Table 2. Comparison of average amplitude and speed $V_s$ of leading pulses taken between the 14$^{th}$ and 27$^{th}$ particles for experiments numerical results in a strongly nonlinear diatomic chain.

|              | Striker Mass [g] | Pulse Speed [m/s] |
|--------------|------------------|-------------------|
| Experimental | 0.61             | 118               |
|              | 1.22             | 121               |
|              | 2.75             | 122               |
|              | 17.81            | 115               |
| Numerical    | 0.61             | 134               |
|              | 1.22             | 141               |
|              | 2.75             | 146               |
|              | 17.81            | 138               |

The mass of the striker in relation to the mass of the cell roughly corresponds to the number of significant pulses observed in experiments. For example, the impact generated by strikers (Al$_2$O$_3$ cylinder) with masses equal to mass of one cell (PTFE sphere and steel cylinder), 2 cells and 4.5 cells generated the one, two and 5 pulses.

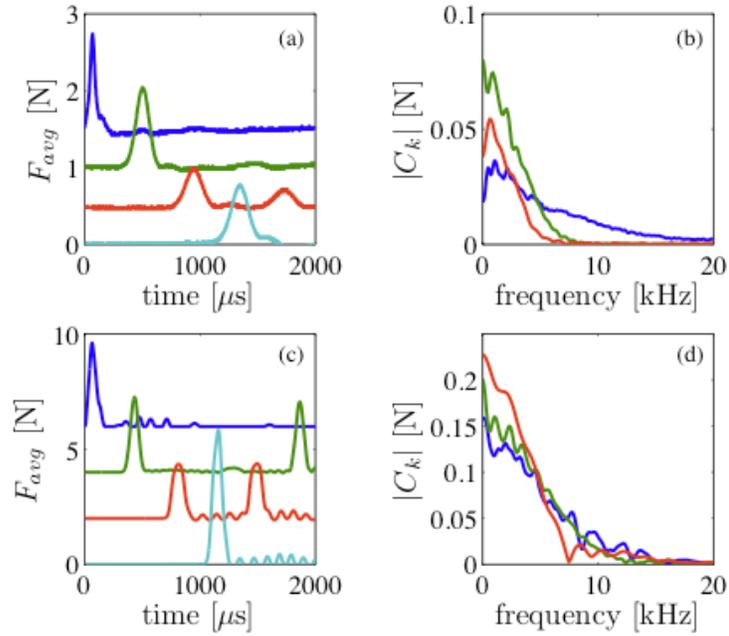

FIG. 11. Strongly nonlinear case with only a gravitational static load. The striker had a mass equal to one cell (0.61 g) and an initial velocity of 0.44 m/s. Experimental data on the propagation of strongly nonlinear solitary waves in a diatomic chain composed of 37 elements. Curves show the force vs. time behavior detected by the sensors in the 2$^{nd}$ particle (top curve), 14$^{th}$ particle (second curve from top), 27$^{th}$ particle and at the wall (bottom curve). The time begins at the moment of impact and is triggered by the first pulse one the oscilloscope.

In Fig. 11 the experimental and numerical results of the impact of a PTFE striker (0.62 g at 0.44 m/s) are shown. This impact created one main pulse in experiments and numerical calculations and the corresponding frequency spectrum of the incoming pulses are shown in Fig. 11 (b) and (d). In Fig. 11 (a), the time duration of the leading solitary wave measured in particle 14 was 330 $\mu$s. The average speed of its propagation between particles 14 and 27 was 118 m/s. These values result in a solitary wave width equal to 4.6 cells. This agrees well with the solitary wave width predicted in the long wave approximation for a chain of particles interacting according to Hertz's law, which predicts a wavelength of approximately five cell lengths [20, 24]. This result corresponds well with numerical results presented in Fig. 11 (c) and Table 2. There are several smaller solitary waves trailing the main pulse apparent in the top curve of Figs. 11 (a) and (c). These waves result from multiple impacts of the first PTFE particle from the top magnetic particle and first stainless steel cylinder.

The Fourier spectrum for the strongly solitary wave and the signal in the nonlinear system (compare Figs. 8(b), (d) with 11(b), (d)) are comparable due to the similar durations of the initial pulses. It should be mentioned that the speed of the linear, weakly nonlinear and strongly nonlinear signals in these cases are different but of the same order of magnitude.

The frequency spectrum shifted toward lower frequencies as the pulse propagates into the chain. This is due to the formation of the solitary wave being a stationary solution [24] of the long-wave approximation with a spectrum frequencies being close to the band gap range presented in the previous section for precompressed chain. This coincidence is due to the specific value of the striker velocity creating amplitude of pulses comparable to the initial precompression in previous cases which results in similar sound speed and solitary wave speeds. A more pronounced shift toward lower frequencies in experiments (Fig. 11(b)) with single pulse is attributed to amplitude attenuation resulting in longer pulse duration with the same spatial width.

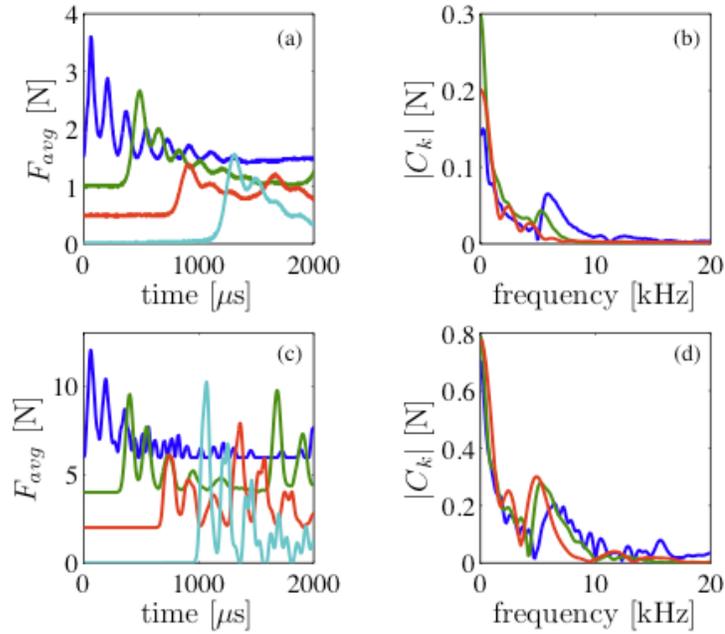

FIG. 12. Strongly nonlinear case, there is no initial compression and only gravity is included. The striker had a mass equal to 4.5 cells (2.75 g) and an initial velocity of 0.44m/s. Experimental data on the propagation of strongly nonlinear solitary waves in a diatomic chain composed of 37 elements. The curves show the force vs. time behavior detected by the sensors in the $2^{nd}$ particle (top curve), $14^{th}$ particle (second from top), $27^{th}$ particle (third from top) and at the wall (bottom curve).

A similar behavior of signals in frequency domain both in experiments and numerical calculations was observed for pulses excited by the impact of an alumina striker with larger mass (1.22 g at 0.44 m/s) generating a series of pulses.

In Fig. 12 the experimental and numerical results of the impact of an alumina striker (2.75 g at 0.44 m/s) are shown. This impact created about 8 pulses in experiments and numerical calculations and the corresponding frequency spectrum of the propagating pulses are shown in Fig. 12 (b) and (d). At the $27^{th}$ (stainless steel cylinder) particle it is clear that the spectrum has shifted to lower frequencies.

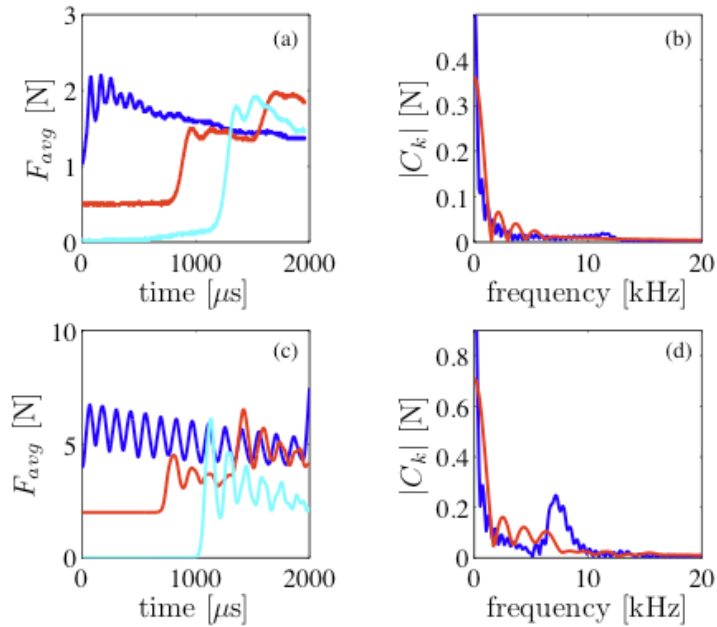

FIG. 13. Oscillatory shock waves in strongly nonlinear case, there is no initial compression and only gravity is included. The striker had a mass equal to 17.81 g and an initial velocity of 0.2 m/s. (a) Experimental data on the propagation of strongly nonlinear solitary waves in a diatomic chain composed of 37 elements (no precompression other than the gravitational preload is included). The Curves show the force vs. time behavior detected by the sensors in the $2^{nd}$ particle (top curve), $27^{th}$ particle (second from the top) and at the wall (bottom curve). (b) Corresponding frequency plots. (c) Numerical calculation corresponding to (a),

(d) Frequency plots obtained in the numerical calculations.

In Fig. 13 the experimental and numerical results of the impact of an alumina striker (17.81 g at 0.2 m/s) are shown. This impact created oscillatory shock waves in experiments and numerical calculations and the corresponding frequency spectrum of the incoming pulses are shown in Figs. 13 (b) and (d). The incoming oscillatory shock becomes almost monotonic by the $27^{th}$ (PTFE) particle in experiments and numerical calculations, which was also observed in case of the nonlinear chain (compare Fig. 10(a), (c) and Fig. 13(a), (c)), despite the fact that dissipation was not included in calculations. The mechanism of changing the oscillatory character of the shock wave without dissipation may be similar to discussed earlier for the case of nonlinear chain. The observed transformation of frequency spectrum is similar to non linear case and may be due to the effects of *in situ* band gap induced by the dynamic compression behind relatively long shock wave. In one mass chain such transformation of oscillatory shock wave in absence of dissipation was not observed [29].

## V. CONCLUSIONS

One dimensional diatomic nonlinear phononic crystals have been assembled and tested using PTFE spheres and steel cylinders. The band gap frequencies can be achieved inside the audible range of frequencies and adjusted by the mass of the particles, their elastic modulus and magnetically induced initial static compression. The transformation of pulses with frequencies inside the phononic band gap is observed in numerical calculations in statically compressed chains. It was shown that the band gap rapidly transforms incoming pulses within very short distances from the source in a linear elastic chain.

It was demonstrated that the band gap calculated in the linear approximation is able to transform of practically harmonic signals with large amplitude causing nonlinear behavior of the system. Initial disturbances separated into two groups of waves; pure acoustic and trailing group of signals with optical frequencies.

Strongly nonlinear solitary waves with a width of five cells were observed in experimental and numerical calculations as predicted by the long wave approximation. The experimental results demonstrate a qualitative change of the propagating wave properties under different conditions of static compression. With the magnetically induced compression, the speed of the solitary wave was increased by 50%.

Oscillatory shock waves were excited under the impact of striker in nonlinear and strongly nonlinear systems. They demonstrated a transition to reduced frequency oscillations both in experiments and calculations without dissipation. This may be explained by *in situ* band gap generated by dynamic compression behind relatively long shock wave. The increased frequency of harmonics in incoming shock profiles in experiments in comparison with numerical calculations may be due to dissipation.

## VI. ACKNOWLEDGEMENTS

This work was supported by NSF (Grant No. DCMS03013220).

[1] J. Tasi, Physical Review B, **14**, 2358 (1976).

[2] St. Pnevmatikos, M. Remoissenet and N. Flytzanis, J. Phys. C: Solid State Phys., **16** (1983).

[3] St. Pnevmatikos, N. Flytzanis, and M. Remoissenet, Phys. Rev. B, 33, 2308 (1986).

[4] O.A. Chubykalo, A.S. Kovalev, and O.V. Usatenko, Phys. Rev. B, **47**, 3153 (1993).

[5] C. Tchawoua, T. C. Kofane and A.S. Bokosah, J. Phys. A: Math. Gen., **26**, 6477 (1993).

[6] S. Parmley, T. Zobrist, T. Clough, A. Perez-Miller, M. Makela, and R. Yu, Appl. Phys. Lett. **67**, 777 (1995).

[7] A. Franchini, V. Bortolani and R.F. Wallis, Phys. Rev. B, **53**, 5420 (1996).

[8] A.V. Gorbach and M. Johansson, Phys. Rev. E, **67**, 066608 (2003).

[9] P. Maniadis, A.V. Zolotaryuk, and G.P. Tsironis, Phys. Rev. E, **67**, 046612 (2003).

[10] D. Lüerßen, N. Easwar, A. Malhotra, L. Hutchins, K. Schulze, and B. Wilcox, Am. J. Phys. **72**, 197 (2004).

[11] C.Kittel, *Introduction to Solid State Physics* (Wiley, 2005), Chap. 4.

[12] A.C. Hladky-Hennion, G. Allan and M. de Billy, J. Applied. Phys, **98**, 054909 (2005).

[13] M. Shen and W. Cao, App. Phys. Lett., **75**, 3713 (1999).

[14] A.T. Alastalo, J. Kiihamaki and H. Seppa, J. of Micromech. Microeng, **16**, 1854 (2006).

[15] V. F. Nesterenko, Prikl. Mekh. Tekh. Fiz. **5**, 136 (1983) [J.Appl. Mech. Tech. Phys. **5**, 733 (1984)].

[16] C. Coste, E. Falcon, and S. Fauve, Phys. Rev. E, **56**, 6104 (1997).

[17] A. Chatterjee, Phys. Rev. E, **59**, 5912 (1999).

[18] E. J. Hinch and S. Saint-Jean, Proc. R. Soc. London, Ser. A, **455**, 3201 (1999).

[19] C. Coste, and B. Gilles, European Physical Journal B, **7**, 155 (1999).

[20] V. F. Nesterenko, *Dynamics of Heterogeneous Materials* (Springer-Verlag, New York, 2001), Chap. 1.

[21] C. Daraio, V. F. Nesterenko, E. Herbold, and S. Jin, Phys. Rev.E, **72**, 016603 (2005).

[22] S. Job, F. Melo, A. Sokolow, and S. Sen, Phys. Rev. Lett., **94**, 178002 (2005).

[23] V.F. Nesterenko, C. Daraio, E.B. Herbold, and S. Jin, Phys. Rev. Lett., **95**, 158702 (2005).

[24] C. Daraio, V. F. Nesterenko, E. Herbold, and S. Jin, Phys. Rev.E, **73**, 026610 (2006).

[25] S. Sen, J. Hong, J. Bang, E. Avalos, and R. Doney, Physics Reports, **462**, 21 (2008).

[26] M.A. Porter, C. Daraio, E.B. Herbold, I. Szelengowicz, P.G. Kevrekidis, Phys. Rev. E, **77**, 015601(R), (2008).

[27] V. Tournat, V.E. Gusev, and B. Castagnede, Phys. Rev. E, **70**, 056603, (2004).

[28] C. Coste, and B. Gilles, Phys. Rev. E, **77**, 021302 (2008).

[29] E.B. Herbold, V.F. Nesterenko, Phys. Rev. E, **75**, 021304 (2007).

[30] G.E. Duvall, R. Manvi, and S.C. Lowell, J. Appl. Phys. **40**, 3771 (1969).

[31] Y. Nahmad-Molinari, C.A. Arancibia-Bulnes, and J.C. Ruiz-Suarez, Phys. Rev. Lett., **82**, 727 (1999).